%% file: iclr2026_conference.tex
\documentclass{article} 
\usepackage{template_files/iclr2026_conference,times}

\input{template_files/math_commands.tex}

\usepackage{hyperref}
\usepackage{url}
\usepackage{graphicx}
\usepackage{booktabs}
\usepackage{enumitem}
\usepackage{array}
\usepackage{authblk}
\title{Helping Customers in Distress: An LLM-powered Agent that Converses, Probes, and Routes}

\author{Alankar Atreya}
\author{Stefan Sylvius Wagner}
\author{Devesh Batra}
\author{Robert Hankache}
\author{Cristovao Iglesias Jr}
\author{Patrick Sinclair}
\author{Giulio Pelosio}
\author{Michael McMillan}
\author{Greig A.\ Cowan}
\author{Raad Khraishi}

\affil{NatWest AI Research}

\iclrfinalcopy 
\begin{document}

\maketitle
\thispagestyle{fancy}

\begin{abstract}
Banks receive millions of reports of fraud, scams, and disputed transactions every year, making it challenging to accurately direct customers to the appropriate specialist teams for assistance. The existing manual process driven by humans is slow and stressful for both customers and staff. To address this, we develop a customer-facing AI powered triaging agent that leverages large language models (LLMs) to conduct multi-turn conversations, ask relevant questions, and classify cases for accurate, policy-guided routing, making it embedded in the customer journey. To evaluate and continuously improve the agent, synthetic digital twins of real customers were simulated, generating realistic, labelled dialogues based on historical data to test a wide range of real-world scenarios. This work details the triage agent’s modelling approach, integration with policy, safety guardrails and reasoning frameworks, the use of the synthetic agent for scalable evaluation, and findings on the AI system’s accuracy, robustness, and compliance. Results show that the agent successfully improves triaging of historical cases, achieving a 30.6\% increase in classification accuracy, with high satisfaction levels reported by our subject-matter experts, highlighting how targeted probing can lead to more effective triage in banking operations at scale.
\end{abstract}

\section{Introduction}
Unauthorised fraud is increasing rapidly in the UK financial sector, rising by over 10\% year-on-year, which is a significant issue for customers and introduces substantial operational challenge to customer service teams \citep{ukfinance_fraud_2025}. Traditional methods of fraud prevention and detection, such as menu-based Interactive Voice Response (IVR) systems and manual triage are becoming unsustainable as case volumes grow. These approaches are often inflexible and lead to misrouted cases and customer dissatisfaction, highlighting the urgent need for scalable, accurate, and compliant automation. Effective triage plays a critical role in this context for customers, it ensures their cases are routed swiftly and accurately, reducing frustration, delays, and repeat contacts; for internal operations, it streamlines workflows by directing cases to the correct teams, reducing manual hand-offs and misdirected workloads, and allowing staff to focus on complex investigations.

To address these issues, we propose a large language model (LLM) powered triage agent layered with safety guardrails that automates the classification and routing of fraud, scam, and dispute cases using a customer-facing chat interface \citep{hou2025llm, batra2025review}. Our solution features advanced prompt engineering \citep{liu2023gpt} to enable rapid adaptation and uses a novel validation framework that links historical case management records with telephony transcripts and supplements this dataset with synthetic “digital twin” customers \citep{sun2024empowering}. These synthetic customers enable the simulation of realistic, multi-turn dialogues at scale, facilitating robust and scalable evaluation of agent accuracy and behaviour in realistic, multi-turn dialogues.

\textbf{Our primary contributions are:} \begin{enumerate}[leftmargin=2em]
\item \textbf{A deployed LLM-based triage agent} that uses multi-turn dialogue to accurately classify fraud, scam, and dispute cases, with integrated handoff protocols and layered safety guardrails for operational reliability.
\item \textbf{A scalable, reproducible evaluation pipeline} that unites operational data and synthetic “digital twin” customers, enabling thorough validation in the absence of labelled chat logs.
\item \textbf{A combined human and automated (LLM) validation strategy} for continuous prompt tuning and quality assurance, demonstrating strong performance, compliance, and readiness for real-world banking deployment.
\end{enumerate}

\section{Methodology}
\begin{figure}[htbp]
    \centering
    \includegraphics[width=0.75\linewidth]{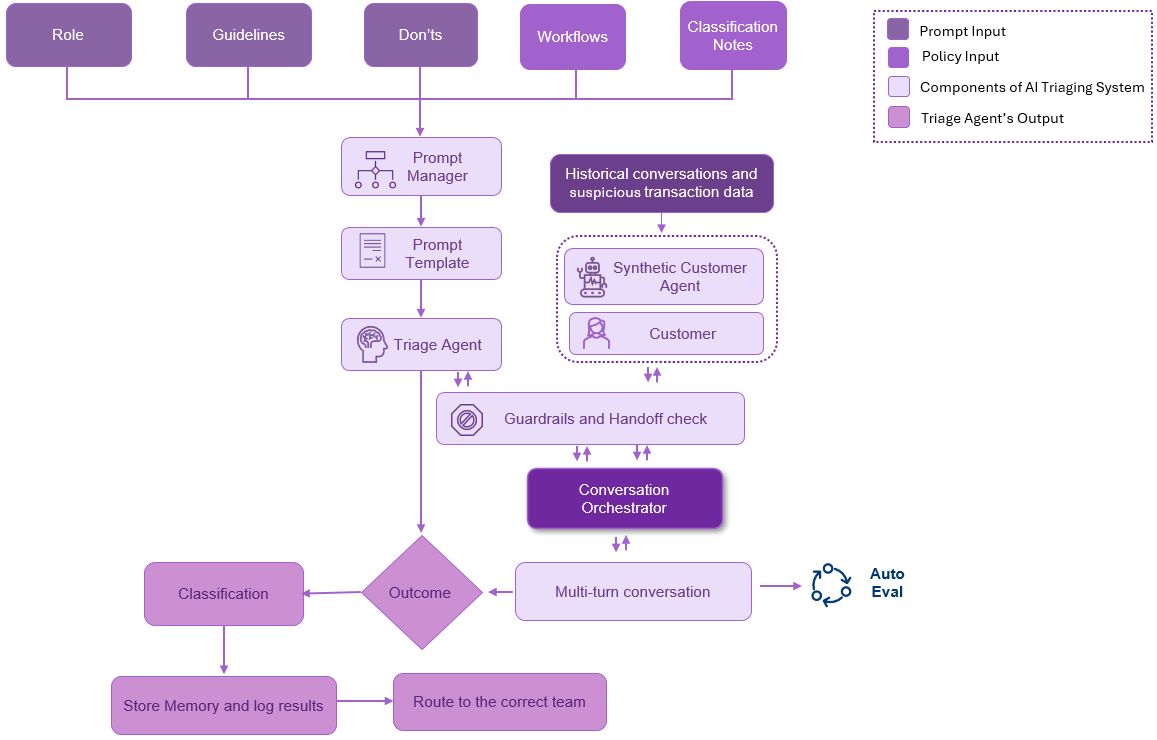}
    \caption{Overview of our triaging framework.}
    \label{fig:overview}
\end{figure}

This section details the design and evaluation of our triage framework, covering the architecture and behaviour of the Triage Agent, handoff and guardrail mechanisms, synthetic “digital twin” customers for large-scale validation, and our combined approach to human and automated performance assessment. An overview is presented in Figure \ref{fig:overview}.

\subsection{Triage Agent}
The Triage Agent handles multi-turn conversations to clarify customer issues and classifies cases as Fraud, Scam, Dispute, or Inconclusive. We utilised third-party LLMs: Claude, Gemini \citep{team2023gemini}, GPT \citep{ai2023gpt}, via prompt engineering, avoiding training or fine-tuning for flexibility, policy compliance, and reduced data risks.
 The triage agent performs \textbf{two tasks:} 
\textbf{(i) multi-turn conversations} where
the agent conducts iterative dialogues, using targeted questions to gather details and summarise key facts. \textbf{(ii) Classification} of the case type, ensuring consistency and leveraging contextual dialogue information.

The Triage Agent’s prompt manager orchestrates specialised sub-prompts that guide conversation and decision-making. These sub-prompts also imbue the Triage Agent with the bank's internal policies. The sub-prompts include \emph{Role}, which defines the agent’s identity along with its goals and duties; \emph{Instructions}, setting behavioural standards emphasising respectful, factual engagement; \emph{Workflow}, outlining the investigative sequence and questioning strategy; \emph{Don’ts}, specifying prohibitions and professional boundaries. Together, these form a structured manual compliant case triage, in accordance with internal policies and external regulations.

\subsection{Hand-offs and Guardrail Agents}

In certain scenarios, it is necessary to end the automated triage chat and transition the customer to an alternative support channel. For example, when the customer expresses the wish to end the conversation or when vulnerability indicators are present. In such cases, an LLM-based digression agent will detect these scenarios and guide the customer through more appropriate routes, such as a direct phone line to a specialist internal team. The agent’s accuracy is measured using a set of conditional trigger phrases, with success rates calculated for each appropriate hand-off.

To ensure safe and compliant operation, the Triage Agent is governed by layered input and output guardrails. AWS Bedrock provides foundational filtering against harmful and manipulative content, supplemented by custom input controls that block unauthorised product requests, probing into internal processes, and non-English messages \citep{bhattacharjee2025introduction}. Effectiveness is benchmarked via an automated pipeline using a test set of thousands of attack prompts and benign examples for input filtering. Output guardrails are validated against a dataset comprising customer messages paired with representative “good” and “bad” responses, quantifying filtering accuracy.

Furthermore, a red-teaming exercise (compliant with internal ethics procedures) was conducted using a production-like web-interface to assess the Triage Agent’s guardrails and ethical robustness. Testers simulated diverse malicious behaviours, such as chat history injection, code manipulation, non-English prompts, and attempts to extract internal reasoning. Identified vulnerabilities were addressed by updating guardrail prompts and input handling as described in the following section.

\subsection{Digital Twins for Large-Scale Validation}

To support scalable validation, we developed transcript-driven ``digital twin'' customer agents. These synthetic agents are designed to mimic real customer behaviour, language, and interaction styles by drawing on historical telephony transcripts. As shown in Figure \ref{fig:overview}, each digital twin is paired with transaction scripts extracted from legacy case data. This ensures the responses reflect genuine scenarios and operational realities.

The digital twins are implemented with LLMs using the OpenAI Agent SDK. The simulation framework uses prompt engineering to automatically generate system prompts by combining original transcripts, customer attributes, and transaction details. This approach enables synthetic dialogues to closely replicate real interactions while adapting dynamically to different scenarios. Furthermore, fidelity tests revealed that synthetic utterances mirrored real customer conversations. 

\subsection{Human-Evaluation and Auto-Evaluation} To robustly assess and improve triage agent performance, we combined large-scale synthetic testing with expert evaluation by subject matter experts (SMEs), ensuring model accuracy, compliance, and real-world reliability. SMEs carried out hands-on evaluation of the Triage Agent using a dedicated web interface, engaging in multi-turn dialogues that reflected genuine customer interaction scenarios. Across three testing rounds, SMEs assessed the agent’s performance using a comprehensive set of metrics: \textbf{satisfaction} (overall user contentment with responses), \textbf{empathy} (sensitivity towards the customer’s situation), \textbf{compliance} (adherence to guidelines and FCA policies), \textbf{factuality} (accuracy of statements such as dates or amounts), \textbf{summary} (fidelity of the closing summary), \textbf{acknowledgement} (recognition of key facts provided by the customer), \textbf{relevancy} (focus on pertinent details), \textbf{language ease} (clarity and accessibility of language), \textbf{frustration} (potential for customer frustration), and \textbf{smoothness} (the logical flow of the conversation). We collected binary (Yes/No) ratings and open-ended comments, enabling a well-rounded view of strengths and areas for improvement.

We developed an automated evaluation pipeline using GPT-4.1 as an “LLM-as-a-judge” \citep{gu2024survey}, which assessed synthetic and real agent-customer conversations, providing binary ratings and rationales. Agreement between human and automated scores was calculated for validation.

\section{Results}
The performance was assessed across three pillars: synthetic scenario testing (using around $3{,}000$ transcript-based cases), SME-led evaluation with feedback obtained from surveys, and handoff triggers and guardrails were systematically stress-tested using both adversarial and benign prompts, ensuring robust protection against manipulation and alignment with operational policy. For synthetic scenario testing and the SME-led evaluation we report results relative to a legacy IVR system.

\begin{table}[h!]
\centering
\caption{Relative overall classification accuracy gains over legacy systems across five large language models.}
\vspace{0.8em}  
\scalebox{0.75}{
\begin{tabular}{lccccc}
\toprule
\textbf{Metric} & \textbf{Claude Sonnet 3} & \textbf{Gemini-1.5-Pro} & \textbf{GPT 4.1-mini} & \textbf{GPT 4.1} & \textbf{GPT 5} \\
\midrule
Synthetic Customer Accuracy Gain & +20.0\% & +28.4\% & +21.3\% & +27.7\% & \textbf{+30.6\%} \\
\ [95\% Confidence Interval] & [+16.0, +23.8] & [+24.8, +31.4] & [+17.5, +25.9] & [+26.5, +33.2] & [+27.1, +33.9] \\
\bottomrule
\end{tabular}}
\label{tab:accuracy_comparison}
\end{table}

SME testing across multiple rounds showed an +16.0\% improvement in average classification accuracy over the legacy baseline. Further integration of SME feedback alongside human and LLM-driven error analysis led to measurable improvements in the Triage Agent's performance. Large-scale evaluation with synthetic customer scenarios allowed for thorough, data-driven assessment of classification accuracy, resulting in consistently high performance gains across models (see Table~\ref{tab:accuracy_comparison}). All models demonstrated robust generalisation against historical case data and diverse scenarios, with GPT 4.1 achieving an overall accuracy improvement of \textbf{+27.7\%}, while preliminary testing with GPT 5 showed similar strong performance with \textbf{+30.6\%} over the legacy baseline. Gemini-1.5-Pro delivered a similar uplift (\textbf{+28.4\%}), followed by GPT 4.1-mini (\textbf{+21.3\%}), and Claude Sonnet 3 (\textbf{+20.0\%}). These strong and reproducible improvements show close alignment between the agent’s predictions and legacy case outcomes, while achieving faster and more efficient resolution across varied synthetic dialogues.

Key evaluation metrics, including compliance, satisfaction, and summary, consistently exceeded the 75–80\% threshold, demonstrating the agent’s operational and conversational reliability as assessed by experts. Internal testing showed strong alignment between automated and human ratings for objective metrics (compliance, factuality, relevance, summary), with agreement from 79\% to 92\%, but notably lower for subjective aspects like empathy and frustration ($\sim 60\%$). Consequently, automated evaluation is dependable for content-focused criteria but less so for subjective ones. Production deployment may require targeted human-in-the-loop validation for emotionally nuanced cases.

Across all key handoff scenarios, the handoff agent achieved precision and recall rates exceeding 90\%, demonstrating consistently high accuracy and reliability in detecting when handoffs are required. These results indicate that the system is both accurate in triggering appropriate handoffs and effective in minimising unnecessary or missed escalations. Guardrail evaluation showed that input guardrails detected prompt injection and hate speech with over 98\% accuracy, while output guardrails identified hallucinatory contents with accuracy exceeding 95\% across both GPT-4.1 variants. Internal red-teaming exercises validated the system’s ability to block nuanced adversarial prompts, underscoring the guardrails’ strength in preventing unauthorised or unsafe content. As a result, the improved guardrail configurations were validated by automated replay of the original attack scenarios, demonstrating effective mitigation of the previously identified vulnerabilities. The system reliably flagged and rejected fabricated conversation histories, enforced English-only interactions, prevented disclosure of internal reasoning strategies, and permitted harmless mentions of protected characteristics and relevant accusations.

In summary, the Triage Agent delivers consistent improvements in classification accuracy over legacy systems, exceeding key thresholds for satisfaction, compliance, and precision in both synthetic and expert-reviewed tests. These results confirm its effectiveness, safety, and suitability for deployment in regulated environments.

\section{Conclusion}
The LLM-based triage agent delivered faster and more efficient resolution of cases, achieving classification accuracy gain of up to \textbf{+30.6\%} over the legacy baseline in synthetic scenario testing. Key operational metrics, including satisfaction, compliance, and accuracy consistently exceeded \textbf{75–80\%}. Automated evaluation closely matched human ratings for objective criteria (agreement rates: \textbf{76–92\%}), with lower agreement for subjective qualities such as empathy and frustration (around \textbf{60\%}). Handoff precision and recall each surpassed \textbf{90\%}; guardrails achieved \textbf{96–98\%} accuracy. These results confirm the agent’s effectiveness, safety, and suitability for high-stakes customer triage in regulated domains.

\subsubsection*{Acknowledgments}
We acknowledge Amanda Miglionico, Daniel Clifton, Darryl Fishwick, and Martin Perry for their domain expertise and insightful discussions on business logic. We also thank Tomoko Komatsu for her support with customer testing, as well as all the subject matter experts (SMEs) who participated in multiple testing rounds and provided valuable feedback for model improvement. Finally, we extend our appreciation to Graham Smith and the Fraud Prevention CoE team for their support and for enabling this work.

\bibliography{references/iclr2026_conference}
\bibliographystyle{template_files/iclr2026_conference}

\end{document}

%% file: template_files/math_commands.tex
\usepackage{amsmath,amsfonts,bm}

\def\eqref#1{equation~\ref{#1}}

\def\1{\bm{1}}

\DeclareMathAlphabet{\mathsfit}{\encodingdefault}{\sfdefault}{m}{sl}
\SetMathAlphabet{\mathsfit}{bold}{\encodingdefault}{\sfdefault}{bx}{n}

